\providecommand{\arxiv}[1]{#1}
\providecommand{\lncs}[1]{}

\arxiv{
\documentclass[12pt]{article}
\usepackage{fullpage}
}
\lncs{
\documentclass[runningheads]{llncs}
\usepackage[T1]{fontenc}
}

\usepackage{hyperref}

\usepackage{alltt}
\usepackage{xspace}

\usepackage{listings}
\usepackage{xcolor}

\usepackage{fancyvrb}
\usepackage{caption}
\newcommand{\Hex}[1]{\hspace{#1ex}}
\newcommand{\Vex}[1]{\vspace{#1ex}}

\newcommand{\mysec}[1]{\Vex{-.75}\section{#1}\Vex{-0.25}}
\newcommand{\mysubsec}[1]{\Vex{-.75}\subsection{#1}\Vex{-0.25}}
\newcommand{\mypar}[1]{\Vex{-1}\paragraph{\bf #1.}}
\newenvironment{code}{\Vex{-1}\begin{alltt}\footnotesize}{\end{alltt}\Vex{-1}}
\NewDocumentCommand{\co}{+m}{\mbox{\footnotesize\tt #1}} %
\newcommand\coline[1]{\Vex{1}\\\co{\Hex{3}#1}\Vex{1}\\} %
\newcommand\m[1]{$#1$} %
\newcommand\p[1]{${\it #1}$} %
\newcommand{\eg}[1]{}
\newcommand{\floor}[1]{\m{\left\lfloor #1 \right\rfloor}}
\newcommand{\bigO}[1]{\m{O(#1)}}
\newcommand{\fig}[1]{\arxiv{ (App.\,A, Fig.\,\ref{#1})}}
\newcommand{\figd}[1]{\arxiv{ (App.\,A, Fig.\,#1)}} %
\newcommand{\fige}[1]{\arxiv{ App.\,A, Fig.\,#1}} %

\newcommand{\todo}[1]{}

\definecolor{codegreen}{rgb}{0,0.6,0}
\definecolor{codegray}{rgb}{0.5,0.5,0.5}
\definecolor{codepurple}{rgb}{0.58,0,0.82}

\makeatletter
\lst@Key{countblanklines}{true}[t]%
    {\lstKV@SetIf{#1}\lst@ifcountblanklines}

\lst@AddToHook{OnEmptyLine}{%
    \lst@ifnumberblanklines\else%
       \lst@ifcountblanklines\else%
         \advance\c@lstnumber-\@ne\relax%
       \fi%
    \fi}
\makeatother

\lstdefinestyle{basestyle}{
  basicstyle=\ttfamily\scriptsize,
  breakatwhitespace=false,         
  breaklines=false, %
  captionpos=b,                    
  keepspaces=true,
  numbers=none, %
  numbersep=5pt,                  
  showspaces=false,                
  showstringspaces=false,
  showtabs=false,
  numberblanklines=false,  %
  belowskip=1ex,
}

\lstdefinestyle{pystyle}{
  style=basestyle, 
  numberstyle=\tiny\color{codegray},
  language=Python,
  commentstyle=\color{codegreen},
  keywordstyle=\color{magenta},
  stringstyle=\color{codepurple},
  basewidth = {.43em} %
}

\lstdefinestyle{dastyle}{
  style=pystyle,
  basicstyle=\ttfamily\tiny,
  morekeywords={main, config, process, new, setup, start, run, self, 
    send, sent, to, receive, msg, received, from_,
    await, timeout, each, some, has, count}
}

\lstdefinestyle{dacodestyle}{
  style=dastyle,
  basicstyle=\ttfamily\scriptsize,
}

\lstdefinestyle{pseudocodestyle}{
  style=basestyle,
  basicstyle=\ttfamily\scriptsize,
  language={},
  numbers=none,
}

\def\src{paxosSB_orig_cleaned.da}

\begin{document}
\arxiv{
\title{Specifying Paxos for System Builders:\\ Pseudocode Made Executable}
\author{Yanhong A. Liu \Hex{12} Rahul Sihag\\
  Stony Brook University, Stony Brook, NY 11733, USA\\
  \{liu,rsihag\}@cs.stonybrook.edu}
\date{}
}

\lncs{
\title{\!\!Invited\,Paper:~Specifying~Paxos~for~System\,Builders:\!\!\\
Pseudocode Made Executable
}
\titlerunning{Invited Paper: Specifying Paxos for System Builders}

\author{Yanhong A. Liu \and Rahul Sihag}

\institute{Stony Brook University, Stony Brook, NY 11733, USA\\
  \email{\{liu,rsihag\}@cs.stonybrook.edu}
}
} %

\maketitle

\begin{abstract}
  This paper presents a precise executable specification---as a faithful
  mapping from the pseudocode---of Paxos for System Builders, a practical
  protocol for replication and consensus in distributed systems.
  Paxos for System Builders has both a robust implementation in C and a
  clean pseudocode for critical protocol details.

  This paper shows how the protocol pseudocode can be expressed %
  easily, essentially line-by-line, in
  a precise high-level language, DistAlgo, for direct execution in
  distributed systems.
  Precise specification and direct execution help significantly in
  understanding the protocol logic and in automatically checking, tracing,
  and visualizing protocol runs.  They also led to discoveries and fixes of
  small, difficult-to-catch omissions and liveness bugs in the pseudocode
  though not the C code.
  The resulting program also has acceptable performance\todo{}
  while having similar size as the pseudocode.
\end{abstract}
\lncs{\Vex{-2}}

\mysec{Introduction}

Paxos for System Builders%
~\cite{kirsch2008paxos,kirsch2008paxosTR} is a complete, practical protocol
for the classical replication and consensus problem in distributed systems.
Replication and consensus protocols are at the core of fault-tolerant
distributed systems, where processes may fail and may later recover, and
messages may be lost, delayed, reordered, and duplicated.  These protocols
are for replicating servers and reaching agreement on the sequence of
client requests served.

Paxos~\cite{Lam98paxos,lam01paxos} is such a protocol, after earlier
protocols Virtual Synchrony~\cite{birman1987reliable,birman1987vs} and
Viewstamped Replication~\cite{oki88vsr,liskov2012vr}. Paxos is well known
for its elaborate story-telling and requirement specification of the
protocol and for presenting different aspects of the protocol separately.
It has been studied extensively, especially to piece the different aspects
together for an overall protocol specification and practical
implementations,
e.g.,~\cite{burrows06chubby,chandra07paxos,kirsch2008paxos,kirsch2008paxosTR,vra15paxos,Liu+19Paxos-PPDP}.

\mypar{Paxos for System Builders (Paxos-SB)}
Among replication and consensus protocols, Paxos-SB~\cite{kirsch2008paxos}
stands out uniquely in not only having a robust practical implementation
but also being both open source~\cite{paxosb} and having a clean protocol
specification with detailed pseudocode explaining critical practical
considerations~\cite{kirsch2008paxosTR}.

Furthermore, Paxos-SB distinguishes itself from Paxos by having what we
call prudent leader election and active normal-case execution, making it
much more robust and responsive in the presence of failures.
\todo{}

The full, working implementation of Paxos-SB~\cite{paxosb} is in C and is
about 7000 lines.\todo{}
The pseudocode including all major protocol components is about 330 lines,
spread in 13 figures~\cite{kirsch2008paxosTR}.

\mypar{This paper}
This paper presents a precise, pseudocode-level, executable specification
of Paxos-SB in DistAlgo~\cite{Liu+12DistPL-OOPSLA,Liu+17DistPL-TOPLAS}, a
high-level language for clear expression and direct execution
of distributed algorithms.
We show how pseudocode can be expressed straightforwardly, basically
line-by-line, as directly executable code in DistAlgo, which requires only
Python to run.

Precise specification and direct execution help significantly in
understanding the protocol logic and in automatically checking, tracing,
and visualizing protocol runs.  They also led to discovery and fixes of
small, %
hard-to-catch omissions and liveness bugs in the pseudocode though not
the C code.
The resulting DistAlgo program is of similar size as the pseudocode,
and about one-tenth as efficient as the C code, with excellent latency.
We discuss the overall lessons learned and directions for future work at
the end.

\arxiv{
The rest of the paper is organized as follows. Section~\ref{sec-background}
gives an overview of Paxos-SB and DistAlgo.
Section~\ref{sec-express} describes expressing Paxos-SB pseudocode as
directly executable code.  Section~\ref{sec-run} presents results from
direct execution, runtime checking, and analysis.
Section~\ref{sec-related} discusses related work and concludes.
Appendix~\ref{app-pseudo} lists the complete Paxos-SB pseudocode
from~\cite{kirsch2008paxosTR}. 
Appendix~\ref{app-orig-da} containts a complete, directly executable specification in DistAlgo.
Appendix~\ref{app-run} shows a run of Paxos-SB in DistAlgo.
}

\mysec{Background}
\label{sec-background}

\mysubsec{Paxos for System Builders}
\label{sec-paxosSB}

Like other classical replication and consensus protocols, Paxos-SB assigns
a global, persistent, total order to operations requested by clients,
called client updates in Paxos-SB.  A server executes an update after it
has executed all previous updates in the global order.

In Paxos and similar 
protocols, a leader is elected at a time by a majority (or quorum) of
servers, who then proposes client updates (each with an assigned sequence
number) to be accepted by a majority (or quorum) of servers.
What distinguishes Paxos-SB from others
include four main aspects:
\begin{enumerate}

\item {\bf Detailed message passing and state tracking}%
  ---detailed sending and receiving of messages with detailed data
  structures and maintenance for state tracking as needed in a complete and
  efficient protocol;

\item {\bf Prudent leader election}---a leader election protocol, which we
  refer to as prudent, 
  for the system to be stable with the current leader in the presence of
  failures that involve only a minority of servers;

\item {\bf Active normal-case execution}---normal-case execution 
  where servers respond to proposals from the leader by sending accepts to
  all servers, not just the leader, so all servers can actively order the
  client updates proposed;
  and

\item {\bf Critical liveness guarantees}---precise characterization and
  comparison of critical liveness
  assumptions and guarantees.

\end{enumerate}
We give an overview of the core protocol, focusing on the first three
aspects above.  Liveness properties for replication and consensus
protocols, especially with Paxos-SB characterizing two different liveness
statements, are discussed
separately~\cite{ChaLiu21Liveness-PODC,ChaLiu20Liveness-arxiv}.

\mypar{Detailed message passing and state tracking}
Paxos-SB uses a full range of types of messages between servers. It uses
extensive data structures in each server to track the server's main state,
various kinds of messages received, ordering status, timers, and client
update handling status.

The main state of a server is one of (1)
\co{LEADER\_ELECTION}---participating in leader election, (2)
\co{REG\_LEADER}---being the
leader that proposes sequence numbers to client updates,
and (3) \co{REG\_NONLEADER}---being a non-leader that forwards client
updates to the leader for sequencing and replies to proposals from the
leader.

As in classical replication and consensus protocols, two key variables are
used in each server, called \co{Last\_Installed} and \co{Local\_Aru} in
Paxos-SB:
\begin{itemize}

\item \co{Last\_Installed}, a.k.a.\ view number: an integer holding the
  installed view number for tracking the protocol progress through successive
  views.
  It is called view in Virtual Synchrony %
  and Viewstamped Replication, %
  and ballot in Paxos. %

\item \co{Local\_Aru}, a.k.a.\ sequence number:
  an integer holding the position in the sequence of
  ordered client updates that have been agreed upon and executed.
  Aru stands for ``All received upto''~\cite{amir1995replication}.

\end{itemize}
Key message types of the protocol are the following 3 pairs:
\begin{itemize}

\item \co{View\_Change} and \co{VC\_Proof}: for informing about progress
  with view, for a newly-attempted view 
  and an already-installed view, %
  respectively.

\item \co{Prepare} and \co{Prepare\_OK}: for the leader of a
  newly-installed view to collect previously ordered or accepted proposals
  from servers, and for servers to reply, respectively.

\item \co{Proposal} and \co{Accept}: for the leader of an installed view to
  propose the next client update in the sequence, and for servers to reply,
  respectively.

\end{itemize}
The 3 pairs of messages are used in the 3 main parts of Paxos-SB: leader
election, prepare phase, and global ordering, respectively, where the
global ordering protocol is the core of normal-case execution.

The data structures, message types, checking of received messages, and data
structure updates upon receiving different types of message are specified
in detail in pseudocode~\cite[Section 5.2]{kirsch2008paxos}\figd{2-5}.

\mypar{Prudent leader election}
Leader election elects a leader among servers to lead normal-case execution
and is crucial for liveness properties of the protocol.  What distinguishes
Paxos-SB most
is its leader election protocol, which we call prudent leader election.
\begin{itemize}

\item 
Paxos and other protocols and variants typically use either a heart-beat or
ping-pong mechanism, with a fixed or random timeout, for a server
to detect whether the leader of the current view has failed to make
progress
and start a new view 
by the timeout.

This has a significant drawback---any server who doesn't hear from the
leader by the timeout unilaterally starts a new view, making the entire
protocol abandon the current view,
even while a majority consisting of other servers are making good progress.

\item Paxos-SB solves this problem with its leader election protocol. It
  uses an extra phase of \co{View\_Change} messages, and only starts a new
  view with a new leader after a majority of servers have timed out and
  decided to move to the new view---called the installed view---with the
  new leader.

  This requires a majority of servers sending \co{View\_Change} messages to
  each other, thus requiring \bigO{N^2} messages, but it ensures that no
  minority of servers can disrupt the system from making progress with a
  majority of servers, an adaption of ideas from PBFT~\cite{castro02PBFT}
  for benign environments.

\end{itemize}
Only after installing the new view, the new leader gets prepared by
collecting previously ordered or accepted proposals from a majority of
servers so as to 
be in agreement with them. 
This collection is as in all classical replication and consensus protocols.

Leader election is specified in detail in Paxos-SB pseudocode~\cite[Section
5.3]{kirsch2008paxosTR}\fig{6-elect}.  If a new view is successfully
installed, i.e., agreed upon by a majority, the leader of the new view
shifts to the prepare phase~\cite[Section
5.4]{kirsch2008paxosTR}\figd{7-8} and, upon success, shifts to the leader
state \figd{9 (block A)}, and other servers
shift to the non-leader state\figd{9 (block B)}.

\mypar{Active normal-case execution}
During normal-case execution, centered around what is called the global
ordering protocol in Paxos-SB, one server is the leader.
\begin{enumerate}
\item[(1)]
  The leader proposes a sequence number to a client update it received, and
  sends a \co{Proposal} message about this to the other servers.
\item[(2)]
  Each non-leader server
  responds to the \co{Proposal} by sending an \co{Accept} message about it
  to all servers.
\item[(3)]
  Each server orders an update, according to a \co{Proposal} it sent or
  received about the proposed sequence number for the update, when it
  receives \floor{N/2} \co{Accept} messages about the proposal.
\end{enumerate}
Note that this is more active than Paxos and other variants, where each
non-leader responds to the \co{Proposal} by sending an \co{Accept} only to
the leader, and thus only the leader can order an update; then the leader
sends the ordering decision to others or the next leader collects the
\co{Accept}'s sent previously. This active execution requires \bigO{N^2}
instead of \bigO{N} messages, but reduces
latency in keeping all servers up-to-date, and is also more fault-tolerant
than counting all on the leader~\cite[Section 4.4]{kirsch2008paxosTR}.

Normal-case execution is specified in detail in the Paxos-SB
pseudocode~\cite[Section 5.5]{kirsch2008paxosTR}\figd{9 (block A)-11},
along with handling of client updates to be proposed~\cite[%
Section 5.7]{kirsch2008paxosTR}\figd{12-13}:
\begin{itemize}

\item It starts with the leader shifting to leader state (using
  \co{Shift\_to\_Reg\_Leader()} in\fige{9}), which tracks prepared updates
  (using
  \co{Enqueue\_Unbound\_Pending\_Updates()} and
  \co{Remove\_Bound\_Updates\_From\_Queue()} in\fige{13}) and then sends
  proposals for ordering the updates received (using \co{Send\_Proposal()}
  in\fige{11}).
\item This continues with receiving \co{Proposal}, then sending and
  receiving \co{Accept}, and finally ordering and executing a
  \co{Client\_Update}, all are after receiving and handling the client
  update (all in\fige{10}).

  Ordering and executing after receiving \co{Accept} checks majority and
  increases \co{Local\_Aru} (using \co{Globally\_Ordered\_Ready(seq)} and
  \co{Advance\_Aru()}, respectively, in\fige{11}).

  Handling client updates after receiving \co{Client\_Update} forwards the
  update to the leader if the server is a non-leader or sends a proposal
  for the update if the server is the leader (using
  \co{Client\_Update\_Handler(U)} in\fige{12}), after bookkeeping (using
  \co{Enqueue\_Update(U)} and \co{Add\_to\_Pending\_Updates(U)} in\fige{13}).
\end{itemize}
\lncs{\Vex{-2}}

\mysubsec{DistAlgo}

For precise executable specification of Paxos-SB at a high level that
corresponds to protocol pseudocode, we use
DistAlgo~\cite{Liu+12DistPL-OOPSLA,Liu+17DistPL-TOPLAS}.
DistAlgo supports four main concepts of distributed programming by
extending an object-oriented programming language, Python.

\mypar{1) Distributed processes that can send messages}
A class \co{\p{P}} of processes is defined by
\begin{code}
    class \p{P} (process): \p{stmt}
\end{code}
where \p{stmt} may contain, among usual definitions,
(i) a \co{setup} definition for setting up the
  values used,
(ii) a \co{run} definition for running the %
  flow of the process\eg{}, and
(iii) \co{receive} definitions for handling received messages\eg{}.
A process refers to itself as \co{self}. Expression \co{self.\p{attr}}
(or \co{\p{attr}} when there is no ambiguity) refers to the value of
\co{\p{attr}}.
To start, \co{\p{ps} = new(\p{P},\,num=\p{n})}
  creates \co{\p{n}} new processes of class \co{\p{P}}
  and assigns them to \co{\p{ps}};
\co{setup(\p{ps},\,\p{args})} sets up \co{\p{ps}} using values of
\co{\p{args}};
and \co{start(\p{ps})} starts \co{run} of \co{\p{ps}}.
Finally, \co{send(\p{m},\,to=\p{ps})} process sends message 
\co{\p{m}} to \co{\p{ps}}.

\mypar{2) Control flow for handling received messages}
Received messages can be handled both asynchronously, using 
\co{receive} definitions, and synchronously, using \co{await}
statements. A \co{receive} definition
\begin{code}
    receive \p{m} from \p{p}: \p{stmt}
\end{code}
handles messages that match \co{\p{m}} from \co{\p{p}} at yield points,
which are implicit before \co{await} statements, 
and executes \p{stmt}.  An \co{await} statement
\begin{code}
    if await \p{cond\sb{1}}:\,\p{stmt\sb{1}} elif\,...\,elif \p{cond\sb{k}}:\,\p{stmt\sb{k}} elif timeout(\p{t}):\,\p{stmt}
\end{code}
waits for one of \co{\p{cond_1}},...,\co{\p{cond_k}} to be true or a
timeout after \co{\p{t}} seconds, and nondeterministically selects one
of \co{\p{stmt_1}},...,\co{\p{stmt_k}}, \co{\p{stmt}} whose conditions
are true to execute.

\mypar{3) High-level queries for synchronization conditions}
High-level queries can be used over message histories, and patterns can be
used to match messages.
\begin{itemize}
  \setlength{\parskip}{.5ex}

\item Messages sent and received by a process are kept in \co{sent} and
  \co{received}, respectively.

\item A pattern can be used to match a message, in \co{sent} and
  \co{received}, and by a \co{receive} definition.  A constant value, such
  as \co{'ack'}, or a previously bound variable, indicated with prefix
  \co{\_}, must match the corresponding components of the message.  An
  underscore \co{\_} matches anything.  Previously unbound variables are
  bound to the corresponding components in the matched message.

  For example, \co{received(('ack',\_t,\_), from\_=p)} matches every
  message triple and sender in \co{received} whose first component is
  \co{'ack'} and second component is the value of \co{t}, ignores the third
  component, and binds \co{p} to the sender.

\end{itemize}
A query can be a universal or existential quantification, a comprehension,
or an aggregation over sets or sequences. For example,
\begin{code}
    each \p{v\sb{1}} in \p{s\sb{1}}, ..., \p{v\sb{k}} in \p{s\sb{k}} has \p{cond}
\end{code}
returns true iff for each combination of values of variables that satisfies
all \co{\p{v\sb{i}} in \p{s\sb{i}}} clauses, \co{\p{cond}} holds\eg{}.  In
all \p{v\sb{i}} in \p{s\sb{i}} forms, \co{\p{v_i}} can be a pattern.

\mypar{4) Configuration for setting up and running}
Configuration for requirements such as use of logical clocks can be
specified in a \co{main} definition.

DistAlgo also supports automatic visualization of replays forward and
backward, making it easier to understand protocol runs.

In our specification, we include extensive comments: \co{\#} indicates
pseudocode or text copied from the Paxos-SB paper~\cite{kirsch2008paxosTR},
\co{\#\#} (or no comments) indicates code we had to fill in, and
\co{\#\#\#} indicates changes to the Paxos-SB pseudocode.

\def\da{

For precise executable specification of Paxos-SB at a high level 
that corresponds to protocol pseudocode,
we use DistAlgo~\cite{Liu+12DistPL-OOPSLA,Liu+17DistPL-TOPLAS}.
DistAlgo is not only high-level, like pseudocode languages, but has also a
formal semantics, like formal specification languages, and is directly
executable, like programming languages.

DistAlgo supports the following four main concepts of distributed 
programming by extending an object-oriented programming language,
Python.
We first present the main concepts using an ideal syntax and then describe
the corresponding Python syntax.

\mypar{1) Distributed processes that can send messages}
A type \co{\p{P}} of processes is defined by
\begin{code}
    process \p{P}: \p{stmt}
\end{code}
The body \p{stmt} may contain, among usual definitions,
\begin{itemize}
\setlength{\itemsep}{0ex}

\item a \co{setup} definition for setting up the
  values used in the process, %

\item a \co{run} definition for running the main %
  flow of the process\eg{}, and

\item \co{receive} definitions for handling received messages\eg{}.

\end{itemize}
A process can refer to itself as \co{self}. Expression \co{self.\p{attr}}
(or \co{\p{attr}} when there is no ambiguity) refers to the value of
\co{\p{attr}} in the process.
\begin{itemize}
\setlength{\itemsep}{0ex}

\item \co{\p{ps} := \p{n} new \p{P}}
  creates \co{\p{n}} new processes of type \co{\p{P}},
  and assigns the new processes to
  \co{\p{ps}}\eg{}.

\item %
\co{\p{ps}.setup(\p{args})} sets up processes \co{\p{ps}} using values of
\co{\p{args}}\eg{}.

\item \co{\p{ps}.start()} starts \co{run()} of \co{\p{ps}}\eg{}.

\end{itemize}
\co{new} can have an additional clause, \co{at \p{node}},
specifying remote nodes where the created processes will run; the default
is the local node.

A process can easily send %
a message \co{\p{m}} to processes \co{\p{ps}}:
\begin{code}
    send \p{m} to \p{ps}
\end{code}\Vex{-1.2}

\mypar{2) Control flow for handling received messages}
Received messages can be handled both asynchronously, using 
\co{receive} definitions, and synchronously, using \co{await}
statements.
\begin{itemize}

\item A \co{receive} definition is of the following form: \coline{receive
    \p{m} from \p{p}:~\p{stmt}} %
  It handles, at yield points, un-handled messages that match \co{\p{m}}
  from \co{\p{p}}\eg{}.
  A yield point is of the form \co{-\,-\,\p{l}}, where \p{l} is a label; it specifies a point in the program where control yields to handling of un-handled messages, if any, and resumes afterwards.
  There is an implicit yield point before each \co{await} statement\eg{}, for handling messages while waiting.
  The \co{from} clause is optional.

\item An \co{await} statement is of the following form: \coline{await
    \p{cond_1}:\,\p{stmt_1} or\,...\,or \p{cond_k}:\,\p{stmt_k}
    timeout\,\p{t}:\,\p{stmt}} %
  It waits for one of \co{\p{cond_1}}, ..., \co{\p{cond_k}} to be true or a
  timeout after period \co{\p{t}}, and then nondeterministically selects
  one of \co{\p{stmt_1}}, ..., \co{\p{stmt_k}}, \co{\p{stmt}} whose
  conditions are true to execute\eg{}.
  Each branch is optional.
  So is the statement in \co{await} with a single branch.
\end{itemize}\Vex{-1.2}

\mypar{3) High-level queries for synchronization conditions}
High-level queries can be used over message histories, and patterns can be
used to match messages.
\begin{itemize}
  \setlength{\parskip}{.5ex}

\item Histories of messages sent and received by a process are kept in
  \co{sent} and \co{received}, respectively.
  \co{sent} is updated at each \co{send} statement, by adding each message
  sent.
  \co{received} is updated at the next yield point if there are un-handled
  messages, by adding un-handled messages before executing all matching
  \co{receive} definitions.

  Expression \co{sent \m{m} to \p{p}} is equivalent to \co{\p{m} to \p{p} in
    sent}.  It returns true iff a message that matches \co{\p{m} to \p{p}}
  is in \co{sent}.
  The \co{to} clause is optional.
  Expression \co{received \m{m} from \p{p}} is similar.

\item A pattern can be used to match a message, in \co{sent} and
  \co{received}, and by a \co{receive} definition.  A constant value, such
  as \co{"release"}, or a previously bound variable, indicated with prefix
  \co{=}, in the pattern must match the corresponding components of the
  message.  An underscore \co{\_} matches anything.  Previously unbound
  variables in the pattern are bound to the corresponding components in the
  matched message.

  For example, \co{received("release",t3,=p2)} %
  matches every triple in \co{received} whose first component is
  \co{"release"} and third component is the value of \co{p2}, and
  binds \co{t3} to the second component.

\end{itemize}
A query can be an existential or universal quantification, a comprehension,
or an aggregation over sets or sequences.
\begin{itemize}

\item An existential quantification and a universal quantification are of
  the following two forms, respectively: \coline{some \p{v\sb{1}} in
    \p{s\sb{1}}, ..., \p{v\sb{k}} in \p{s\sb{k}} has \p{cond}}\Vex{-5}
  \coline{each \p{v\sb{1}} in \p{s\sb{1}}, ..., \p{v\sb{k}} in \p{s\sb{k}}
    has \p{cond}} %
  They return true iff for some or each, respectively, combination of
  values of variables that satisfies all \co{\p{v\sb{i}} in \p{s\sb{i}}}
  clauses, \co{\p{cond}} holds\eg{}.

\item A comprehension is of the following form:
  \coline{\{\p{e}:~\p{v\sb{1}} in \p{s\sb{1}}, ..., \p{v\sb{k}} in
    \p{s\sb{k}}, \p{cond}\}} %
  It returns the set of values of \co{\p{e}} for all combinations of values
  of variables that satisfy all \co{\p{v_i} in \p{s\sb{i}}} clauses and
  condition \co{\p{cond}}\eg{}.

\item An aggregation is of the form \co{\p{agg} \p{s}}, where \co{\p{agg}}
  is an aggregation operator such as \co{count} or \co{max}.  It returns
  the value of applying \co{\p{agg}} to the set value of \co{s}\eg{}.

\item In all query forms above, each \co{\p{v_i}} can be a pattern.

\end{itemize}
Other operations, such as set union and sequence concatenation, can also be used.

\mypar{4) Configuration for setting up and running}
Configuration for requirements such as the use of logical clocks and the use of
reliable and FIFO channels can be specified in a \co{main} definition.
For example, \co{configure channel = fifo} specifies that fifo channels are used and TCP is used for process communication. 

DistAlgo also supports automatic visualization of replays forward and backward, making it much easier to understand protocol runs.

\mypar{DistAlgo compiler and Python syntax} %
To allow anyone with Python to run DistAlgo directly,
DistAlgo compiler supports the Python syntax~\cite{Distalgo17lang}.
For example, \co{send m to p} is written as \co{send(m, to=p)}, and
\co{each sent m to p has cond} is written as \co{each(sent(m, to=p),}
\co{has=cond)}; in patterns, \co{=var} is written as \co{\_var}.

In our specification in DistAlgo, we include the corresponding pseudocode
as comments.
The following convention for comments (green) are used:
(1) comments after \co{\#} are pseudocode or text
copied from the Paxos-SB paper~\cite{kirsch2008paxosTR},
(2) comments after \co{\#\#} (or no comments) describe code we had to fill in; and
(3) comments after \co{\#\#\#} describe changes to the Paxos-SB pseudocode.
\todo{}
} %

\mysec{From pseudocode to directly executable code}
\label{sec-express}

Paxos-SB pseudocode captures a highly nontrivial and critical protocol
design in detail.  We show how its
pseudocode can be expressed straightforwardly, essentially with
line-by-line correspondence, in DistAlgo.

We focus on two core parts of the protocol: leader election with progress
timers and \co{View\_Change} and \co{VC\_Proof} messages, and global
ordering with \co{Proposal} and \co{Accept} messages.
We skip other 
parts, where the use of DistAlgo is even more straightforward: the prepare
phase with \co{Prepare} and \co{Prepare\_OK} messages, client update
handling with queue operations and update timers, etc.

\mysubsec{Data structures and message types}

The pseudocode first defines the data structures %
and message types %
used,
with checks on received messages to filter out undesired messages %
and with updates to data structure %
based on the type of messages received.

All data structures defined in the pseudocode\fig{2-data}, for data
maintained in each server, are scalars, arrays, structures, or queues,
including nested.  They are represented directly using built-ins in
DistAlgo as in Python, using object for structure, and list for array and
queue, except that, when efficient lookups using more general keys are
needed,
dictionary is used instead of list.

For example,
below are 3 of 6 parts
of the data structures in the pseudocode, %
for variables related to the main state, view, and, global ordering history that each
server maintains. 
\begin{code}\scriptsize
    /* Server State variables */
    A1. int My\_server\_id - a unique identifier for this server
    A2. State - one of {LEADER\_ELECTION, REG\_LEADER, REG\_NONLEADER} \medskip
    /* View State variables */
    B1. int Last\_Attempted - the last view this server attempted to install
    B2. int Last\_Installed - the last view this server installed
    B3. VC[] - array of View Change messages, indexed by server id \medskip
    /* Global Ordering variables */
    D1. int Local\_Aru - the local aru value of this server
    D2. int Last\_Proposed - last sequence number proposed by the leader
    D3. Global_History[] - array of global slots, indexed by sequence number, each containing:
    D4. Proposal - latest Proposal accepted for this sequence number, if any
    D5. Accepts[] - array of corresponding Accept messages, indexed by server id
    D6. Globally\_Ordered\_Update - ordered update for this sequence number, if any
\end{code}
They are directly variables in the process class for servers in DistAlgo,
as follows. 
\co{global\_history} (last in \co{setup})
keeps the history sequence of globally ordered updates, each with the
proposal and acceptances for the update.  The structure for each element of
\co{global\_history} is defined separately using class
\co{Global\_History}.
As in Python, \co{\_\_init\_\_} constructs a class instance, \co{self}, 
with its fields initialized using the argument values.
\co{\_\_repr\_\_} returns the string representation of \co{self} when it
is printed in output.\todo{}
\lstinputlisting[style=dacodestyle,linerange={86-88}]{\src}
\lncs{\Vex{-1}}
\lstinputlisting[style=dacodestyle,linerange={92-94}]{\src} %
\lncs{\Vex{-1}}
\lstinputlisting[style=dacodestyle,linerange={96-99}]{\src} %
\lncs{\Vex{-1}}
\lstinputlisting[style=dacodestyle,linerange={106-112}]{\src} %
\lncs{\Vex{-1}}
\lstinputlisting[style=dacodestyle,linerange={78-84}]{\src}
Note that initial values are not specified in the pseudocode except
for 5 scalar variables in the recovery procedure,
but we created all and confirmed them with the C code.
\co{global\_history} is represented using a dictionary to allow efficient
lookups when older history need not be held in memory; in fact
\co{Global\_History[]} in the pseudocode is implemented using a hash table in
the C code.

Message types\fig{3-msg} are %
expressed as classes, e.g., for message types
\co{View\_Change}, %
\co{Proposal}, and \co{Accept}
below, out of 8 types total.
\lstinputlisting[style=dacodestyle,linerange={23-27}]{\src}
\lncs{\Vex{-1}}
\lstinputlisting[style=dacodestyle,linerange={49-55}]{\src}
\lncs{\Vex{-1}}
\lstinputlisting[style=dacodestyle,linerange={57-62}]{\src}
Note that each message has a field \co{server\_id} indicating the sending
server of the message. It could be omitted when expressed in DistAlgo and,
in the few places where it is needed, we can simply use the server in the
\co{from\_} clause in the received message, but we keep it just to be close
to the pseudocode.

Checks on received messages\fig{4-check} are checks on the messages using
their fields, e.g., for \co{View\_Change}\arxiv{, \co{Proposal},} and \co{Accept}
messages below, in a method \co{Conflict}, exactly as in the
pseudocode.\todo{}
\co{State} is a class object encapsulating the three main states.
\co{Progress\_Timer\_} is the timer for progress in a view.
\lstinputlisting[style=dacodestyle,linerange={143,145-154}]{\src} %
\arxiv{
\lstinputlisting[style=dacodestyle,linerange={177-184}]{\src}%
}
\lncs{\Vex{-1}}
\lstinputlisting[style=dacodestyle,linerange={186-195}]{\src}%

Updates for data structures\fig{5-update} maintain the data in variables.
For the data structures defined earlier in this section, they are updated
when \co{View\_Change}, \co{Proposal}, and \co{Accept} messages are
received, and the updates are to \co{VC} and \co{global\_history}, as
follows.
\lstinputlisting[style=dacodestyle,linerange={199-204}]{\src}%
\lncs{\Vex{-1}}
\lstinputlisting[style=dacodestyle,linerange={216-247}]{\src}%

In all code blocks, we can see that the DistAlgo code corresponds to the
pseudocode exactly, line-by-line, other than a few lines for defining a
class, printing, or similar. The only exception is an added update to an
added data structure \co{bound\_updates} after line D7 and an added
base-case check and initialization in the last \co{elif} branch. The former
fixed an omission that \co{bound\_updates} is used but not defined, as
discussed in Section~\ref{sec-missing}.

\mysubsec{Leader election}
\label{sec-elect}

The protocol
critically depends on leader election, to elect a leader at a time, who
then coordinates the ordering of client updates.

As for Paxos-SB pseudocode in general, the leader election protocol has
several blocks of code.  Some are triggered upon received messages or timer
timeouts, and some are functions or procedures that are called explicitly.
Each block can be expressed directly as a method definition, and those
triggered by messages or timeouts are called when the message is received
or the timeout occurs. 
However,
\begin{itemize}
\item Those triggered by received messages can also be directly expressed
  with a \co{receive} handler, not needing a method definition that is
  called in a \co{receive} handler.
\item Those triggered by timer timeouts can also be expressed with a
  \co{receive} handler for a timeout message sent, instead of a method
  called, when the timeout occurs.
\end{itemize}
Calling methods is the more traditional procedural programming model, and
we use it when it matches the Paxos-SB pseudocode more exactly.  Otherwise,
we simply use \co{receive} handlers.

For the leader election protocol specifically\fig{6-elect}, there are 5
blocks, and they are expressed as 5 methods in DistAlgo, as 
below. %
The first method is called when progress timer expires. The second and
third are called when receiving a \co{View\_Change} or \co{VC\_Proof}
message, respectively.  The fourth
is called in only the second block (line B7), and the fifth is called in
only the first two blocks (lines A2 and B3) and \co{run}, in the entire
Paxos-SB pseudocode.
\todo{}
\lstinputlisting[style=dacodestyle,linerange={258-303}]{\src}%
\todo{}  

Calls to the methods for receiving \co{View\_Change} and \co{VC\_Proof},
among all types of messages that require checks using \co{Conflict}, are in
a \co{receive} handler, as below. They are put together in a single
\co{receive}, so that \co{Conflict} is only called here before further
handling using different methods for different message types.
\lstinputlisting[style=dacodestyle,linerange={539-545}]{\src}
\todo{}

The key ideas for leader election using \co{View\_Change} messages in each
server are as follows, achieving what we call prudent leader election,
especially through point (2):
\begin{itemize}
\item[(1)] when \co{Progress\_Timer\_} expires (line A1), or when a
  \co{View\_Change} message is received with an larger attempted view and
  the progress time already expired (lines B1-2),
  \co{Shift\_to\_Leader\_Election} is called (lines A2 and B3), which sets
  the state and related data structures for leader election and sends a
  \co{View\_Change} message with a newly attempted view (lines E1-5); and
\item[(2)] only when receiving \co{View\_Change} messages with the same
  attempted view from a majority of servers (lines B7 and D1-5), is the
  server willing to move to a new leader---the designated leader of the
  newly attempted view---and if the server, \co{self}, is the leader of
  this attempted view (line B10), it moves to the prepare phase (line B11).
\end{itemize}
The prepare phase then lets the new leader collect previously accepted
proposals from a majority of servers, so that new proposals agree with
them, ensuring consensus.

Each server also periodically sends \co{VC\_Proof} messages with its
installed view to other servers.
When a \co{VC\_Proof} message with a larger installed view is received
(lines C1-2), the server takes the view in the message as the newly
attempted view (line C3) and moves to the prepare phase if it is the new
leader (lines C4-5) or to be a non-leader otherwise (lines C6-7).

We can see that the DistAlgo code corresponds to the pseudocode exactly,
line-by-line, with two small coding exceptions: (1) a timer object and its timeout
value are in two variables in DistAlgo, instead of one (line B8) in the
pseudocode, and (2) clearing data structures for 4 variables uses 4 lines in
DistAlgo instead of one line (line E2) in the pseudocode.

There are also two  exceptions for the logic in \co{Shift\_to\_Leader\_Election}: (1)
at the start, an update to \co{state} is added (after line E1), and (2) at
the end, a line was added before the last \co{send} (line E5), and the last
line (line E6) was removed. The first maintains the correct state needed in
many places in the protocol.  The second fixed a liveness bug in the
pseudocode, as discussed in Section~\ref{sec-bugs}.

\mysubsec{Global ordering}
\label{sec-order}

After the prepare phase for a new view, the leader can propose received
client updates for global ordering.\todo{}

The global ordering protocol has 4 main blocks\fig{10-order}, and they are
expressed in DistAlgo as below.  The first block is for receiving a
\co{Client\_Update} message, expressed directly as a \co{receive}
definition.  The second and third blocks are invoked when receiving a
\co{Proposal} or \co{Accept} message, respectively.  The fourth block is
for executing a client update that has just been ordered.
\lstinputlisting[style=dacodestyle,linerange={378-418}]{\src}%
There are also 3 utility procedures\fig{11-order-util} called in the main
blocks, and they are expressed in DistAlgo as below.  The first,
\co{Send\_Proposal()}, for sending a proposal for the next client update if
any, is called by the leader at the end of prepare phase, 
at the end of executing a \co{Client\_Update} here, and when receiving a
new client update.
The second, \co{Globally\_Ordered\_ready(seq)}, for checking having received a
majority of \co{Accept} messages for a \co{Proposal}, is called when
receiving an \co{Accept} message.
The third, \co{Advance\_Aru()}, is called when \co{Globally\_Ordered\_ready(seq)}
returns true.\todo{}
\lstinputlisting[style=dacodestyle,linerange={421-450,451-461}]{\src}%

The key ideas for global ordering in each server are as follows, achieving
what we call active normal-case execution, especially through point (2); we
refer to the 4 main blocks as ``main'' and 3 utility blocks as ``util'':
\begin{itemize}
\item[(1)] when receiving a \co{Client\_Update} message, call
  \co{Client\_Update\_Handler} (lines A1-2 in main), which
  adds the update to \co{Pending\_Updates} if it is a new update from a
  client of the server, and forwards the update to the leader if the state is
  \co{REG\_NONLEADER}, or calls \co{Send\_Proposal()} to propose the next
  update to all servers %
  if the state is \co{REG\_LEADER};

\item[(2)] when receiving a \co{Proposal} for an update from a leader, send
  an \co{Accept} for it to all servers (lines B1-5 in main), not only to
  the leader, thus allowing all servers to react actively;
  and for any server, not just the leader, when receiving \co{Accept}s from
  one fewer than a majority of servers (lines C3 in main and B1-5 in util),
  record a successfully ordered update (lines C4-5 in main) and call
  \co{Advance\_Aru()} (lines C6 in main and C1-8 in util) to execute each
  next successfully ordered update in order.
\end{itemize}
Client handling includes \co{Client\_Update\_Handler} explained in (1) and
\co{Update\_Timer} to resend a client update to the leader if the update
has not been executed before the timer expires.

We again see that the DistAlgo code corresponds to the pseudocode
line-by-line, with small exceptions, similar to \co{Leader\_Election}, such
as for the progress timer (line D10 in main).

A few other exceptions are small fixes: %
(1) for receiving \co{Proposal}, a line was added before
\co{sync\_to\_disk} and sending \co{Accept} at the end (lines B4-5 in
main);
(2) in \arxiv{\linebreak}\co{Upon executing a Client\_Update}, the call to \co{Advance\_Aru()}
(line D2 in main) was removed, and
in the definition of \co{Advance\_Aru()}, after incrementing \co{Local\_Aru}
(line C5 in util), a call to \co{Upon executing a Client\_Update} is
added.
The first fixed a small liveness bug as discussed in Section~\ref{sec-bugs}.
The second fixed an omission that \co{Upon executing a Client\_Update} is
never invoked in the pseudocode, while \co{Advance\_Aru()} is called
unnecessarily in \linebreak 
\co{Upon executing a Client\_Update}, as discussed in
Section~\ref{sec-missing}

\mysubsec{Complete executable specification}

Besides the protocol logic in the pseudocode, a complete executable
specification or implementation also needs \co{import} of time-related
libraries, definitions of time-related parameters, small helper functions
omitted in the pseudocode, a definition of \co{Client} processes, and
\co{main} for creating and setting up all processes and starting the
execution.
These are all straightforward to write in DistAlgo, and the resulting
specification can be directly executed, as described in
Section~\ref{sec-run}.

Note that the recovery procedure is not implemented separately, for clarity
of the data structures and initializations.  Recovery only requires a
library call to write some variables to disk, in \co{sync\_to\_disk()}, and
small changes in the \co{setup} method to initialize those variables with
values read from disk,
instead of empty or default initial values, 
This %
affects the program size and execution minimally
but clutters the specification of data structures and initializations.
Also, not writing to disk is the default for Paxos-SB in C, and is what we
use for performance comparison.

\mysec{Direct execution, checking, and analysis}
\label{sec-run}

Detailed Paxos-SB pseudocode has allowed us to create a precise
specification easily and execute it directly.  This further allows
automated safety and liveness checking as well as tracing and visualization
of protocol runs, which helped significantly in understanding the protocol.
These also allowed us to discover and fix small, difficult-to-catch
omissions and liveness bugs in the pseudocode.
We describe main results from direct execution, checking, and analysis.
\lncs{More detailed description of the omissions and liveness bugs fixed
  can be found in an extended version~\cite{LiuSih26PaxosSB-arxiv}}.

Runtime checking of safety and liveness properties uses a general method in
DistAlgo~\cite{LiuSto20DistCheck-RV}, as used for many protocols, e.g.,
Derecho~\cite{shi+23DerechoDA-ApPLIED}.  With the method, properties can be
specified as high-level queries, and process and communication failures can
be specified by
killing, stopping, and restarting processes and overriding the \co{send}
and \co{receive} methods, without changing the protocol code.  For
Paxos-SB, usual agreement and validity properties were specified for safety,
and various timeouts were specified for liveness.  The liveness bugs in the
pseudocode were discovered by simply killing one of three servers at the
start.

Note that the omissions and bugs are not in the C code. It is a well-known
challenge to keep the design and implementation in sync.
Given that the pseudocode spans 13 figures, with hundreds of lines of
intertwined logic organized manually, it is remarkable how few issues had
to be addressed.
The fixes have all been confirmed with the Paxos-SB team.  Some fixes were
also confirmed first with the C code, but confirmation with the Paxos-SB
team was important when our code is different from the C code and has
different trade-offs.

\mysubsec{Omissions fixed} %
\label{sec-missing}

Two small but necessary pieces were missing in the pseudocode that
prevented a complete execution of the DistAlgo code.

\mypar{Executing a client update is not invoked}
Functionality for executing a client update\arxiv{\!}
\begin{code}
  \arxiv{  }Upon executing a Client\_Update(client\_id, server\_id, timestamp, update), U
\end{code}
is defined in the pseudocode but is never invoked.  It is described as: ``A
server executes an update with an assigned sequence number when its
\co{Local\_Aru} reaches that sequence number.''\,\cite[Section
5.5]{kirsch2008paxosTR}.

\arxiv{
\textbf{\textit{Fix.}}
}
\co{Local\_Aru} is incremented in \co{Advance\_Aru()}, in a \co{while}-loop
over entries of \co{Global\_History}. So it seems obvious to just add an
invocation of this functionality at each such increment.  But this
functionality starts with calling \co{Advance\_Aru()}, meaning a
complicated mutual recursion would be formed.  The fix is to also remove
this call to \co{Advance\_Aru()}, so \co{Advance\_Aru()} is only called
when receiving enough \co{Accept}'s.

\mypar{Checking whether a client update is bound is not defined}
Boolean function \co{bound(U)} is called in two client update handling
functions in the pseudocode but is not defined.  The term ``bound'' is
introduced as: ``once an update is bound to a sequence number, we prevent
the leader from binding the update to any other sequence number (unless it
learns that the update could not have been ordered with the first sequence
number).''\,\cite[Section 5.7]{kirsch2008paxosTR}.

The two calls are with other conditions tracking updates being ordered by
the leader, and the two enclosing functions are both called when a server
becomes a leader, the very core of Paxos.  A new leader must take the most
up-to-date data---by taking, for each sequence number \co{seq}, the update
assigned \co{seq} in a \co{Proposal} with the highest view.  \co{bound(U)}
checks whether \co{U} is such an update.

\arxiv{
\textbf{\textit{Fix.}}
}
Our fix is to (1) define a mapping \co{bound\_updates} from sequence number
\co{seq} to the corresponding client update assigned to \co{seq}; (2) when
processing a \co{Proposal} for a client update \co{U} and with a higher
view, record \co{U} in \co{bound\_updates}; and (3) define \co{bound(U)} to
return whether \co{U} is recorded.

This fix is different from the C code, which just uses a loop that checks
updates at each call to \co{bound(U)}.
When the number of bound updates is large, this is slow because the loop
takes linear time.  Our fix does just a set and a check, in a small
constant time, but uses extra space for \co{bound\_updates}.

\mysubsec{Liveness bugs fixed}
\label{sec-bugs}

When checking executions of DistAlgo programs with failed servers and
detecting some servers becoming stuck, we discovered two liveness bugs in
the pseudocode.

\mypar{Liveness bug due to waiting for \co{Accept}}

The protocol in the pseudocode may become stuck with a server waiting for
enough \co{Accept}'s for a \co{Proposal}.  Paxos-SB counts the sending or
receiving of a \co{Proposal} as a vote and thus a server waits for
\co{Accept}'s from one fewer than a majority of servers, as in
\co{Globally\_Ordered\_Ready(seq)}.
However, in the pseudocode, a server never receives and counts the
\co{Accept} from itself and thus may become stuck.

For efficiency, Paxos-SB uses IP multicast to broadcast messages to all
servers.  In case a message should be sent to all other servers excluding
self, Paxos-SB checks and drops the message from self when the message is
received, in function \co{Conflict}; in particular, \co{Accept} messages
from self are dropped.

\arxiv{
Consider one of 3 given servers crashed, leaving 2 servers, say \co{p0} and
\co{p1}, where \co{p0} is elected leader.
\begin{enumerate}
\setlength{\itemsep}{0ex}

\item \co{p0} sends a \co{Proposal} for an update to \co{p0} and \co{p1};

\item \co{p1} receives the \co{Proposal} and sends an \co{Accept} to
  \co{p0} and \co{p1};

\item \co{p0} receives an \co{Accept} from \co{p1}, and counting its sent
  \co{Proposal}, 1+1 votes, orders the proposed update as expected.

  \co{p1}, however, having received the \co{Proposal} and sent the
  \co{Accept} to all, drops and never counts the \co{Accept} from itself.
  With only 1 vote, \co{p1} never orders the proposed update.

\end{enumerate}
So in this situation, while \co{p1} forwards requested updates from its
clients to \co{p0}, to be proposed and ordered, itself never orders them
and never replies to its clients about them.
}

\arxiv{
\textbf{\textit{Fix.}}
}
A seemingly obvious fix is to not drop \co{Accept} messages from self in
\co{Conflict}, which was what we first did, but this is not sufficient
because a \co{Accept} to self might be dropped by the network before
reaching \co{Conflict} check.
A proper fix, as in the C code and pointed out by the Paxos-SB team, is
instead to let the \co{Accept} be received by self first, i.e., add an
invocation of \co{Upon receiving Accept} on the server's own \co{Accept}
before broadcasting the \co{Accept}.

\mypar{Liveness bug due to waiting for \co{View\_Change}}

This is similar to the liveness bug due to waiting for \co{Accept}, except
it can make all servers and the entire protocol stuck.
This is because, similar to \co{Accept} messages, \co{View\_Change} messages
from self are also dropped in \co{Conflict}, and \co{View\_Change} messages
can be sent by multiple servers at the same time.

\arxiv{
Consider one of 3 given servers crashed, leaving 2 servers, and there
is a temporary network-wide delay.
\begin{enumerate}
\setlength{\itemsep}{0ex}

\item Both servers' progress timers for the current \co{view} expire.

\item Both send a \co{View\_Change} for \co{view + 1} to both.

\item Both receives the \co{View\_Change} from the other server, but never
  receives the one from itself. With only 1 vote, both become stuck waiting
  for one more \co{View\_Change} to successfully install \co{view + 1}.

\end{enumerate}
Unlike the bug due to waiting for \co{Accept}, where \co{p0} can still make
progress, here both servers and the entire protocol become stuck.
}

\arxiv{
\textbf{\textit{Fix.}}
}
The fix is similar to the fix for \co{Accept} except to also remove\lncs{\linebreak}
\co{Apply vc to data structures} after broadcasting \co{View\_Change},
because that is a very partial attempt to achieve the effect of the added
invocation.

\def\fixed{
\mysubsec{Omissions fixed} %
With pseudocode completely expressed in DistAlgo, two small but necessary
pieces were missing in the pseudocode that prevented a complete execution
of the DistAlgo code.

\mypar{Executing a client update is not invoked}
Functionality for executing a client update
\arxiv{
\begin{code}
    Upon executing a Client\_Update(client\_id, server\_id, timestamp, update), U
\end{code}
}
\lncs{
\begin{code}
  Upon executing a Client\_Update(client\_id, server\_id, timestamp, update), U
\end{code}
}
is defined in the pseudocode but is never invoked.  It is described as: ``A
server executes an update with an assigned sequence number when its
\co{Local\_Aru} reaches that sequence number.''\,\cite[Section
5.5]{kirsch2008paxosTR}.

\textbf{\textit{Fix.}}
\co{Local\_Aru} is incremented in function \co{Advance\_Aru()}, even
repeatedly in a \co{while}-loop going through entries of
\co{Global\_History} (Section~\ref{sec-order}, line C5 in util).  So our
fix was to add an invocation of this functionality in \co{Advance\_Aru()}
for each such incremented value of \co{Local\_Aru}.

This addition would have been easy and sufficient 
except that, in the pseudocode, this 
functionality calls \co{Advance\_Aru()} as the first statement
(Section~\ref{sec-order}, line D2 in main).  Thus, the addition would
create a complicated mutual recursion that is also intertwined with a
\co{while}-loop.

We then determined that this call of \co{Advance\_Aru()} in the pseudocode
should be removed, and it is sufficient to leave in the only call to it
just after receiving enough \co{Accept}'s from servers to form a majority.
So our fix also includes this removal.

With both the addition and removal, we have that, after receiving enough
\co{Accept}'s for a \co{Proposal}, a server calls \co{Advance\_Aru()}
(Section~\ref{sec-order}, line C6 in main), which iteratively increments
\co{Local\_Aru} through each next sequence number that has been assigned a
client update, and invokes the functionality for executing the update
(Section~\ref{sec-order}, lines C1-8 in util).

\mypar{Checking whether a client update is bound
  is not defined}

Function \co{bound(U)} is called in two places in the pseudocode but is not
defined.  The term ``bound'' is introduced when describing client handling:
``once an update is bound to a sequence number, we prevent the leader from
binding the update to any other sequence number (unless it learns that the
update could not have been ordered with the first sequence
number).''\,\cite[Section 5.7]{kirsch2008paxosTR}.

The function returns a Boolean. 
It is called in client update handling utility functions
\co{Enqueue\_Unbound\_Pending\_Updates()} and
\co{Remove\_Bound\_Updates\_From\_Queue()}\fig{13-client-util}, together
with other conditions on several data structures tracking client updates
being or to be ordered by the leader.  Both of those functions are called
in \co{Shift\_to\_Reg\_Leader()} when a server just became a leader after
receiving \co{Prepare\_OK} from a majority of servers.

This is at the very core of Paxos, where a new leader must ensure that it
takes the most up-to-date information from other servers---by taking, for
each sequence number \co{seq} for which a update is yet to be decided, the
update assigned \co{seq} in a \co{Proposal} with the highest prior view.
\co{bound(U)} checks whether \co{U} is such an update.

\textbf{\textit{Fix.}}
  Our fix has three parts: (1) in the data structures,
  define a mapping \co{bound\_updates}, mapping a sequence number \co{seq}
  to the corresponding client update assigned to \co{seq}; (2) in
  \co{Update\_Data\_Structures(message)}, when processing a \co{Proposal}
  for a client update and with a higher view,
  record the update in \co{bound\_updates};
  and (3) define \co{bound(U)} to return whether update \co{U} is recorded
  in \co{bound\_updates}.\todo{}
  
  This fix is different from what is in the C code, where no data structure
  is used to record bound updates, but a loop is used to do the check at
  calls to \co{bound(U)}.
  When the number of bound updates is large, this is slow because looping
  through them takes linear time.  Our fix does just a set and a check,
  taking a small constant time, and thus is faster, but uses extra space
  storing \co{bound\_updates}.

\mysubsec{Liveness bugs fixed}
When checking executions of the DistAlgo program with failed servers and
observing some servers becoming stuck, not making progress, we discovered
two bugs in the pseudocode that may cause liveness violations.

\mypar{Liveness bug due to waiting for \co{Accept}}

The protocol in the pseudocode may become stuck with a server waiting for
enough \co{Accept}'s for a \co{Proposal}.  Paxos-SB counts the sending or
receiving of a \co{Proposal} as a vote and thus a server waits for
\co{Accept}'s from one fewer than a majority of servers, as defined in
\co{Globally\_Ordered\_Ready(seq)}\,in Section~\ref{sec-order}.
However, in the pseudocode, a server never receives and counts the
\co{Accept} from itself and thus may become stuck with not making progress
in ordering client updates.

For efficiency, Paxos-SB uses IP multicast to broadcast messages to all
servers.  In case a message should be sent to all other servers excluding
self, Paxos-SB achieves this by checking and dropping the message from self
when the message is received, as defined in function
\co{Conflict}\cite[Section 5.2]{kirsch2008paxosTR}.  This is more efficient
than sending separately to every other server but self.
In particular, \co{Accept} messages from self are dropped as defined in
\co{Conflict}.

Consider one of 3 given servers crashed, leaving 2 servers, say \co{p0} and
\co{p1}, where \co{p0} is elected leader.
\begin{enumerate}
\setlength{\itemsep}{0ex}

\item \co{p0} sends a \co{Proposal} for an update to \co{p0} and \co{p1};

\item \co{p1} receives the \co{Proposal} and sends an \co{Accept} to
  \co{p0} and \co{p1};

\item \co{p0} receives an \co{Accept} from \co{p1}, and counting its sent
  \co{Proposal}, 1+1 votes, orders the proposed update as expected.

  \co{p1}, however, having received the \co{Proposal} and sent the
  \co{Accept} to all, drops and never counts the \co{Accept} from itself.
  With only 1 vote, \co{p1} never orders the proposed update.

\end{enumerate}
So in this situation, while \co{p1} forwards requested updates from its
clients to \co{p0}, to be proposed and ordered, itself never orders them
and never replies to its clients about them.

\textbf{\textit{Fix.}}
A seemingly obvious fix may be to not check and drop \co{Accept} messages
from self when receiving the message, which was what we first did.
However,
this could create an inconsistency---and in fact the same bad situation as
just shown above---if an \co{Accept} fails to reach self but is received by
other servers.

The correct fix, as done in the C code and pointed out by the Paxos-SB
team, is instead to let the \co{Accept} be received by self first, i.e.,
add an invocation of \co{Upon receiving Accept} on the server's own
\co{Accept} just before \co{sync\_to\_disk} and sending the \co{Accept}
(Section~\ref{sec-order}, lines B4-5 in main).
So overall, 4 steps are performed: (1) for the \co{Accept} to be sent in
response to a \co{Proposal}, invoke \co{Upon receiving Accept} on this
\co{Accept}, which does \co{Apply Accept to data structures} and checks and
possibly orders and executes the client update in the \co{Proposal}; (2) do
\co{sync\_to\_disk}; (3) send the \co{Accept} to all servers; and (4) when
receiving an \co{Accept}, check and drop it if is sent from self, as
defined in \co{Conflict}.

\mypar{Liveness bug due to waiting for \co{View\_Change}}

This is similar to the liveness bug due to waiting for \co{Accept}, except
it can make all servers and the entire protocol stuck.

Similar to \co{Accept} messages, \co{View\_Change} messages from self are
also dropped as defined in \co{Conflict}.
Consider one of 3 given servers crashed, leaving 2 servers, and there
is a temporary network-wide delay.
\begin{enumerate}
\setlength{\itemsep}{0ex}

\item Both servers' progress timers for the current \co{view} expire.

\item Both send a \co{View\_Change} for \co{view + 1} to both.

\item Both receives the \co{View\_Change} from the other server, but never
  receives the one from itself. With only 1 vote, both become stuck waiting
  for one more \co{View\_Change} to successfully install \co{view + 1}.

\end{enumerate}
Unlike the bug due to waiting for \co{Accept}, where \co{p0} can still make
progress, here both servers and the entire protocol become stuck.

\textbf{\textit{Fix.}}
The fix is similar to the fix for waiting for \co{Accept} except with an
added removal: (1) add an invocation of \co{Upon receiving View\_Change} on
the server's own \co{View\_Change} just before sending it
(Section~\ref{sec-elect}, lines E5), so 3 steps similar to sending and
receiving \co{Accept} but just without \co{sync\_to\_disk} are performed;
and (2) remove \co{Apply vc to data structures} just after the sending of
\co{View\_Change} to all, because it is a very partial and inadequate
attempt to achieve the effect of (1) and is now done in full extend by the
added invocation.
} %

\mysubsec{Resulting specification and direct execution} %

Table~\ref{tab-loc} summarizes the size of the specification of Paxos-SB in
DistAlgo, including the complete protocol specification,
client process definition, and all configurations and setups needed to 
run.  It is less than 600 lines total, but about 370 lines without
about 80 comment-only lines,
85 blank lines separating classes, methods, and blocks,
and 65 lines for client processes and overall setups.
It is interesting to note that the core protocol logic---leader election,
prepare phase, and global ordering---are only about 200 lines total.
The complete resulting formal executable specification in DistAlgo is shown
in \arxiv{Appendix~\ref{app-orig-da}}\lncs{an extended version~\cite{LiuSih26PaxosSB-arxiv}}.

\begin{table}[t]
\centering
\lncs{%
\scriptsize
}
\caption{Lines of code including all comments and blank lines for
  readability.}
\begin{tabular}{l | r}
  Protocol component & Size\\\hline
  Message types and data structures & 124\\ %
  Conflict checks on received messages and updates to data structures & 115\\\hline
  Leader election & 48\\
  Prepare phase   & 72\\
  Global ordering & 87\\\hline
  Client handling & 59\\
  \co{import}s, parameters, \co{run}, helpers, and choices in \co{receive} & 49\\
  Client process definition and main definition for execution & 44\\
  \hline
  Total & 598\\\hline
\end{tabular}\Vex{-2}
\label{tab-loc}
\end{table}

DistAlgo programs can be run directly using Python with or without separate
installation.  For example, if the DistAlgo program is in file
\co{paxosSB.da}, then it can be run directly with Python 3.7
by just running \co{pip install pyDistAlgo} to install DistAlgo and then
running \co{python -m da paxosSB.da}. Later Python versions can run with
slightly different ways.
The output of a run %
is shown in \arxiv{Appendix~\ref{app-run}}\lncs{an extended version~\cite{LiuSih26PaxosSB-arxiv}}.

The performance of the DistAlgo program is quite reasonable even though
DistAlgo is currently compiled to Python, and well-written Python programs
can be 10-100 times than C~\cite{lion2022investigating,van2025pl_perform}.
For example, running 3 servers with 2 clients and 100,000 requests each
client takes less than 170 seconds, averaged over 5 runs.
The average latency of each request is less than 1.7 ms.
The running times are elapsed times measured on an Intel Core Ultra 7
Processor 258V (base frequency 2.2 GHz, max turbo frequency 4.8 GHz) with
32GB RAM, running Windows 11, Python 3.9.13 and DistAlgo 1.2.0b2.
It is about 10 times the running time of the C code.
Indeed, an earlier version of the specification in DistAlgo is about 100
times slower than the C code, from using extra tuples over objects for
messages and so on, even though it lacks some robust handling of boundary
cases that are prone to faults. That was before we performed systematic
runtime checking of safety and liveness, finding and fixing the issues
discovered, and optimizing out the extras.

\mysec{Related work and conclusion}
\label{sec-related}

There have been significant efforts on specification, analysis, and
verification of different replication and consensus protocols,
e.g.,~\cite{hawblitzel2015ironfleet,Cha+16PaxosTLAPS-FM,tholoniat2022formal},
especially as these protocols are critical in distributed systems.
There are also been many specification languages and verification tools
developed to support these efforts,
e.g.,~\cite{Lam02book,Liu+17DistPL-TOPLAS,pirlea2025veil}, for
understanding, checking, and proving such complex protocols.
We chose Paxos for System Builders because it is meticulously designed and
implemented for practical fault tolerance and high performance.
We choose DistAlgo because it is closest to pseudocode languages yet has a
formal semantics and is directly executable.

Many efforts focus on formal specifications and proofs. For example, the
IronFleet project~\cite{hawblitzel2015ironfleet} develops in Dafny a large
specification and mechanically checked proofs for safety and liveness
properties of a Paxos-based state machine replication system.
The Verdi work~\cite{wilcox2015verdi}, built on Coq, develops a Raft
variant specification and
a mechanically checked safety proof.
Chand et al~\cite{Cha+16PaxosTLAPS-FM} develops a specification in TLA+ and
a mechanically checked proof in TLAPS for safety properties of Multi-Paxos.
Padon et al~\cite{padon2017paxos} develops tailored specifications and
encodings of additional properties for Paxos and variants in a restricted
language in Ivy for automated
proofs.
With Veil~\cite{pirlea2025veil}, built using Lean, similar specifications
and proofs as in Ivy have been developed.
These specifications are either large and complex,
e.g.,~\cite{hawblitzel2015ironfleet,wilcox2015verdi}, or omit many
important practical aspects,\lncs{\linebreak}
e.g.,~\cite{Cha+16PaxosTLAPS-FM,padon2017paxos,pirlea2025veil}.
In contrast, our specification in DistAlgo is mapped almost exactly from
the pseudocode designed by experts in practical distributed systems.

There are drastically more works on model checking of consensus
protocols~\cite{feng2026survey}, not to mention runtime verification and
testing for correctness and performance,
e.g.,~\cite{chandra07paxos,ailijiang2019dissecting,bano2022twins},
especially for practical systems\arxiv{, not just abstract specifications}.
Practical systems in lower-level languages are generally orders of
magnitude more complex than pseudocode, making them drastically more
challenging to understand, let alone verify or improve.  We use runtime
checking for DistAlgo~\cite{LiuSto20DistCheck-RV} because it allows
complete implementations to be checked directly at the level of pseudocode
for easier, faster, and more precise understanding, and it has proven
effective, e.g., for runtime checking~\cite{shi+23DerechoDA-ApPLIED} of
Derecho~\cite{derecho2019,jha2022rdma}.

Many protocols have been written easily in
DistAlgo~\cite{LiuSto24DistConsensus-ICDCStut}, and runtime verification
and testing has helped find and fix not only programming bugs but often
protocol issues,
e.g.,~\cite{Liu+19Paxos-PPDP,shi+23DerechoDA-ApPLIED,Bod25thesis}.
No prior work has developed precise formal specifications of Paxos-SB, in
any language or tool, not just DistAlgo.  Additionally, no prior work on
formal proofs or model checking for Paxos and variants reported finding any
issues and fixes or improvements, except for using DistAlgo on Moderately
Complex Paxos~\cite{Liu+19Paxos-PPDP}.
However, moderately complex Paxos~\cite{vra15paxos} is still much simpler
than Paxos-SB, missing all main aspects described in
Section~\ref{sec-paxosSB}.
Also, it lets the leader decide only views and lets replicas decide sequence
numbers, creating unnecessary inefficiencies and liveness
issues~\cite{ChaLiu21Liveness-PODC}.

Some may find that Fonseca et al.~\cite{fonseca2017empirical} discovered
over a dozen bugs in IronFleet~\cite{hawblitzel2015ironfleet},
Verdi~\cite{wilcox2015verdi}, and Chapar~\cite{lesani2016chapar}, where
IronFleet and Verdi specified variants of Paxos.  However, one should note
that while IronFleet, Verdi, and Chapar developed formal specifications and
proofs of distributed consensus and key-value store protocols, none of them
reported finding any bugs, fixes, or improvements in the original
descriptions or implementations of the actual protocols.  What Fonseca et
al.~\cite{fonseca2017empirical} discovered are bugs in those formal
specifications and proofs, not bugs in the original descriptions of
implementations of the actual protocols.

Our conclusion is that, overall, expressing protocol pseudocode in DistAlgo
is straightforward, and precise executable specifications have helped us
understand the protocol precisely and furthermore find and fix omissions
and liveness bugs in the pseudocode.
As ongoing and future work, our Paxos-SB specification is helping
tremendously in developing an even higher-level, declarative specification,
similar to~\cite{Liu+19Paxos-PPDP} but for the much more practical and
sophisticated Paxos-SB, to allow even easier protocol understanding and
better improvements.

\Vex{-1.25}
\subsection*{Acknowledgment}\Vex{-.75}
We thanks the Paxos-SB team, for their remarkably detailed pseudocode, and
Yair Amir and Amy Babay in particular, for all their helps in answering our
questions, about not only Paxos-SB's implementation and experiments but
also design choices and trade-offs.
Thanks also to Uns Rehman, Neha Mane, Vaibhav Rustagi, Amy Babay, and Scott
Stoller for additional testing and evaluation on various versions of our
specifications.
This work was supported in part by NSF under grant CCF-1954837.

{
\def\usebib{
\def\bibdir{../../bib}   %
{
\bibliography{\bibdir/strings,\bibdir/liu,%
\bibdir/Sec,\bibdir/Veri,\bibdir/Sys,\bibdir/Perform,\bibdir/SE,\bibdir/Vis,\bibdir/misc,\bibdir/crossref} %

\arxiv{\bibliographystyle{alpha}}
\lncs{\bibliographystyle{splncs04}}
}
}%
\usebib
}

\appendix

\mysec{Paxos for System Builders pseudocode}
\label{app-pseudo}
This contains all figures for the pseudocode of Paxos-SB.  The figures are
numbered the same as those for the original Paxos-SB pseudocode~\cite{kirsch2008paxosTR}.
\setcounter{figure}{1}

\begin{figure}[htbp]
  \centering
\scriptsize
\begin{Verbatim}[commandchars=\\\{\},codes={\catcode`$=3},frame=single,baselinestretch=1]
{\bf{}Data Structures:}
/* Server State variables */
A1. int My_server_id - a unique identifier for this server
A2. State - one of {LEADER_ELECTION, REG_LEADER, REG_NONLEADER}

/* View State variables */
B1. int Last_Attempted - the last view this server attempted to install
B2. int Last_Installed - the last view this server installed
B3. VC[] - array of View_Change messages, indexed by server_id

/* Prepare Phase variables */
C1. Prepare - the Prepare message from last preinstalled view, if received
C2. Prepare_oks[] - array of Prepare_OK messages received, indexed by server_id

/* Global Ordering variables */
D1. int Local_Aru - the local aru value of this server
D2. int Last_Proposed - last sequence number proposed by the leader
D3. Global_History[] - array of global_slots, indexed by sequence number, each containing:
D4.  Proposal - latest Proposal accepted for this sequence number, if any
D5.  Accepts[] - array of corresponding Accept messages, indexed by server_id
D6.  Globally_Ordered_Update - ordered update for this sequence number, if any

/* Timers variables */
E1. Progress_Timer - timeout on making global progress
E2. Update_Timer - timeout on globally ordering a specific update

/* Client Handling variables */
F1. Update_Queue - queue of Client_Update messages
F2. Last_Executed[] - array of timestamps, indexed by client_id
F3. Last_Enqueued[] - array of timestamps, indexed by client_id
F4. Pending_Updates[] - array of Client_Update messages, indexed by client_id
\end{Verbatim}
  \caption{Data Structures}
  \label{2-data}
\end{figure}

\begin{figure}[htbp]
  \centering
\scriptsize
\begin{Verbatim}[frame=single,baselinestretch=.85]
Message Types:
A1. Client_Update - contains the following fields:                     
A2.   client_id - unique identifier of the sending client
A3.   server_id - unique identifier of this client's server
A4.   timestamp - client sequence number for this update
A5.   update - the update being initiated by the client

B1. View_Change - contains the following fields:
B2.   server_id - unique identifier of the sending server
B3.   attempted - view number this server is trying to install

C1. VC_Proof - contains the following fields:
C2.   server_id - unique identifier of the sending server
C3.   installed - last view number this server installed

D1. Prepare - contains the following fields:
D2.   server_id - unique identifier of the sending server
D3.   view - the view number being prepared
D4.   local_aru - the local aru value of the leader

E1. Prepare_OK - contains the following fields:
E2.   server_id - unique identifier of the sending server
E3.   view - the view number for which this message applies
E4.   data_list - list of Proposals and Globally_Ordered_Updates

F1. Proposal - contains the following fields:
F2.   server_id - unique identifier of the sending server
F3.   view - the view in which this proposal is being made
F4.   seq - the sequence number of this proposal
F5.   update - the client update being bound to seq in this proposal

G1. Accept - contains the following fields:
G2.   server_id - unique identifier of the sending server
G3.   view - the view for which this message applies
G4.   seq - the sequence number of the associated Proposal

H1. Globally_Ordered_Update - contains the following fields:
H2.   server_id - unique identifier of the sending server
H3.   seq - the sequence number of the update that was ordered
H4.   update - the client update bound to seq and globally ordered
\end{Verbatim}
  \caption{Message Types}
  \label{3-msg}
\end{figure}

\begin{figure}[htbp]
  \centering
\scriptsize
\begin{Verbatim}[commandchars=\\\{\},codes={\catcode`$=3},frame=single,baselinestretch=.85]
boolean Conflict(message):
   case message:
A1.   View_Change VC(server_id, attempted):
A2.      if server_id = My_server_id
A3.         return TRUE
A4.      if State $\neq$ LEADER_ELECTION
A5.         return TRUE
A6.      if Progress_Timer is set
A7.         return TRUE
A8.      if attempted $\leq$ Last_Installed
A9.         return TRUE
A10.  return FALSE

B1.   VC_Proof V(server_id, installed):
B2.      if server_id = My_server_id
B3.         return TRUE
B4.      if State $\neq$ LEADER_ELECTION
B5.         return TRUE
B6.   return FALSE

C1.   Prepare(server_id, view, leader_aru):
C2.      if server_id = My_server_id
C3.         return TRUE
C4.      if view $\neq$ Last_Attempted
C5.         return TRUE
C6.   return FALSE

D1.   Prepare_OK(server_id, view, data_list):
D2.      if State $\neq$ LEADER_ELECTION
D3.         return TRUE
D4.      if view $\neq$ Last_Attempted
D5.         return TRUE
D6.   return FALSE

E1.   Proposal(server_id, view, seq, update):
E2.      if server_id = My_server_id
E3.         return TRUE
E4.      if State $\neq$ reg nonleader
E5.         return TRUE
E6.      if view $\neq$ Last_Installed
E7.      return TRUE
E8.      return FALSE

F1.   Accept(server_id, view, seq):
F2.      if server_id = My_server_id
F3.         return TRUEF
4.      if view $\neq$ Last_Installed
F5.         return TRUE
F6.      if Global_History[seq] does not contain a Proposal from view
F7.         return TRUE
F8.   return FALSE
\end{Verbatim}
  \caption{Conflict checks to run on incoming messages. Messages for which a conflict exists are discarded.}
  \label{4-check}
\end{figure}

\begin{figure}[htbp]
  \centering
\scriptsize
\begin{Verbatim}[commandchars=\\\{\},codes={\catcode`$=3},frame=single,baselinestretch=.85]
Update_Data_Structures(message):
   case message:
A1.   View_Change V(server_id, view):
A2.      if VC[server_id] is not empty
A3.         ignore V
A4.      VC[server_id] $\leftarrow$ V

B1.   Prepare P(server_id, view, leader_aru):
B2.      Prepare $\leftarrow$ P

C1.   Prepare_OK P(server_id, view, data_list):
C2.      if Prepare_OK[server_id] is not empty
C3.         ignore P
C4.      Prepare_OK[server_id] $\leftarrow$ P
C5.      for each entry e in data_list
C6.         Apply e to data structures

D1.   Proposal P(server_id, view, seq, update):
D2.      if Global_History[seq].Globally_Ordered_Update is not empty
D3.         ignore Proposal
D4.      if Global_History[seq].Proposal contains a Proposal P'
D5.         if P.view > P'.view
D6.            Global_History[seq].Proposal $\leftarrow$ P
D7.            Clear out Global_History[seq].Accepts[]
D8.      else
D9.         Global_History[seq].Proposal $\leftarrow$ P

E1.   Accept A(server_id, view, seq):
E2.      if Global_History[seq].Globally_Ordered_Update is not empty
E3.         ignore A
E4.      if Global_History[seq].Accepts already contains \floor{N/2} Accept messages
E5.         ignore A
E6.      if Global_History[seq].Accepts[server_id] is not empty
E7.         ignore A
E8.      Global_History[seq].Accepts[server_id] $\leftarrow$ A

F1.   Globally_Ordered_Update G(server_id, seq, update):
F2.      if Global_History[seq] does not contain a Globally_Ordered_Update
F3.         Global_History[seq] $\leftarrow$ G
\end{Verbatim}
  \caption{Rules for updating the Global\_History.}
  \label{5-update}
\end{figure}

\begin{figure}[htbp]
  \centering
\scriptsize
\begin{Verbatim}[commandchars=\\\{\},codes={\catcode`$=3},frame=single,baselinestretch=.85]
Leader_Election():
A1. Upon expiration of Progress_Timer:
A2.   Shift_to_Leader_Election(Last_Attempted+1)

B1. Upon receiving View_Change(server_id, attempted) message, V:
B2.   if attempted > Last_Attempted and Progress_Timer not set
B3.      Shift_to_Leader_Election(attempted)
B4.      Apply V to data structures
B5.   if attempted = Last_Attempted
B6.      Apply V to data structures
B7.      if Preinstall_Ready(attempted)
B8.         Progress_Timer $\leftarrow$ Progress_Timer*2
B9.         Set Progress_Timer
B10.        if leader of Last_Attempted
B11.           Shift_to_Prepare_Phase()

C1. Upon receiving VC_Proof(server_id, installed) message, V:
C2.   if installed > Last_Installed
C3.      Last_Attempted $\leftarrow$ installed
C4.      if leader of Last_Attempted
C5.         Shift_to_Prepare_Phase()
C6.      else
C7.         Shift_to_Reg_Non_Leader()

D1. bool Preinstall Ready(int view):
D2.   if VC[] contains \floor{N/2} + 1 entries, v, with v.attempt = view
D3.      return TRUE
D4.   else
D5.      return FALSE

E1. Shift_to_Leader_Election(int view):
E2.   Clear data structures: VC[], Prepare, Prepare_oks, Last_Enqueued[]
E3.   Last_Attempted $\leftarrow$ view
E4.   vc $\leftarrow$ Construct_VC(Last_Attempted)
E5.   SEND to all servers: vc
E6.   Apply vc to data structures
\end{Verbatim}
  \caption{Leader Election}
  \label{6-elect}
\end{figure}

\begin{figure}[htbp]
  \centering
\scriptsize
\begin{Verbatim}[commandchars=\\\{\},codes={\catcode`$=3},frame=single,baselinestretch=.85]
A1. Shift_To_Prepare_Phase()
A2.   Last_Installed $\leftarrow$ Last_Attempted
A3.   prepare $\leftarrow$ Construct_Prepare(Last_Installed, Local_Aru)
A4.   Apply prepare to data structures
A5.   data_list $\leftarrow$ Construct_DataList(Local_Aru)
A6.   prepare_ok $\leftarrow$ Construct_Prepare_OK(Last_Installed, data_list)
A7.   Prepare_OK[My_server_id] $\leftarrow$ prepare_ok
A8.   Clear Last_Enqueued[]
A9.   **Sync to disk
A10.  SEND to all servers: prepare

B1. Upon receiving Prepare(server_id, view, aru)
B2.   if State = LEADER_ELECTION /* Install the view */
B3.      Apply Prepare to data structures
B4.      data_list $\leftarrow$ Construct_DataList(aru)
B5.      prepare_ok $\leftarrow$ Construct_Prepare_OK(view, data_list)
B6.      Prepare_OK[My_server_id] $\leftarrow$ prepare_ok
B7.      Shift_to_Reg_Non_Leader()
B8.      SEND to leader: prepare_ok
B9.   else /* Already installed the view */
B10.     SEND to leader: Prepare_OK[My_server_id]

C1. Upon receiving Prepare_OK(server_id, view, data_list)
C2.   Apply to data structures
C3.   if View_Prepared_Ready(view)
C4.      Shift_to_Reg_Leader()
\end{Verbatim}
  \caption{Prepare Phase}
  \label{7-prep}
\end{figure}

\begin{figure}[htbp]
  \centering
\scriptsize
\begin{Verbatim}[commandchars=\\\{\},codes={\catcode`$=3},frame=single,baselinestretch=.85]
A1. datalist_t Construct_DataList(int aru)
A2.   datalist $\leftarrow \emptyset$
A3.   for each sequence number i, i > aru, where Global_History[i] is not empty
A4.      if Global_History[i].Ordered contains a Globally_Ordered_Update, G
A5.         datalist $\leftarrow$ datalist $\cup$ G
A6.      else
A7.         datalist $\leftarrow$ datalist $\cup$ Global_History[i].Proposal
A8.   return datalist

B1. bool View_Prepared_Ready(int view)
B2.   if Prepare_oks[] contains \floor{N/2} + 1 entries, p, with p.view = view
B3.      return TRUE
B4.   else
B5.      return FALSE
\end{Verbatim}
  \caption{Prepare Phase Utility Functions}
  \label{8-prep-util}
\end{figure}

\begin{figure}[htbp]
  \centering
\scriptsize
\begin{Verbatim}[commandchars=\\\{\},codes={\catcode`$=3},frame=single,baselinestretch=.85]
A1. Shift_to_Reg_Leader()
A2.   State $\leftarrow$ REG_LEADER
A3.   Enqueue_Unbound_Pending_Updates()
A4.   Remove_Bound_Updates_From_Queue()
A5.   Last_Proposed $\leftarrow$ Local_Aru
A6.   Send_Proposal()
B1. Shift_to_Reg_Non_Leader()
B2.   State $\leftarrow$ REG_NONLEADER
B3.   Last_Installed $\leftarrow$ Last_Attempted
B4.   Clear Update_Queue
B5.   **Sync to disk
\end{Verbatim}
  \caption{Shift\_to\_Reg\_Leader() and Shift\_to\_Reg\_Non\_Leader() Functions}
  \label{9-leader-nonleader}
\end{figure}

\begin{figure}[htbp]
  \centering
\scriptsize
\begin{Verbatim}[commandchars=\\\{\},codes={\catcode`$=3},frame=single,baselinestretch=.85]
Global Ordering Protocol:
A1. Upon receiving Client_Update(client_id, server_id, timestamp, update), U:
A2.   Client_Update_Handler(U)

B1. Upon receiving Proposal(server_id, view, seq, update):
B2.   Apply Proposal to data structures
B3.   accept $\leftarrow$ Construct_Accept(My_server_id, view, seq)
B4.   **Sync to disk
B5.   SEND to all servers: accept

C1. Upon receiving Accept(server_id, view, seq):
C2.   Apply Accept to data structures
C3.   if Globally_Ordered_Ready(seq)
C4.      globally_ordered_update $\leftarrow$ Construct_Globally_Ordered_Update(seq)
C5.      Apply globally_ordered_update to data structures
C6.      Advance Aru()

D1. Upon executing a Client_Update(client_id, server_id, timestamp, update), U:
D2.   Advance_Aru()
D3.   if server_id = My_server_id
D4.      Reply to client
D5.      if U is in Pending_Updates[client_id]
D6.         Cancel Update_Timer(client_id)
D7.         Remove U from Pending_Updates[]
D8.   Last_Executed[client_id] $\leftarrow$ timestamp
D9.   if State $\neq$ LEADER_ELECTION
D10.     Restart Progress_Timer
D11.  if State = REG_LEADER
D12.     Send_Proposal()
\end{Verbatim}
  \caption{Global Ordering Protocol}
  \label{10-order}
\end{figure}

\begin{figure}[htbp]
  \centering
\scriptsize
\begin{Verbatim}[commandchars=\\\{\},codes={\catcode`$=3},frame=single,baselinestretch=.85]
A1. Send_Proposal()
A2.   seq $\leftarrow$ Last_Proposed + 1
A3.   if Global_History[seq].Globally_Ordered_Update is not empty
A4.      Last_Proposed++
A5.      Send_Proposal()
A6.   if Global_History[seq].Proposal contains a Proposal P
A7.      u $\leftarrow$ P.update
A8.   else if Update_Queue is empty
A9.      return
A10.  else
A11.     u $\leftarrow$ Update_Queue.pop()
A12.  proposal $\leftarrow$ Construct_Proposal(My_server_id, view, seq, u)
A13.  Apply proposal to data structures
A14.  Last_Proposed $\leftarrow$ seq
A15.  **Sync to disk
A16.  SEND to all servers: proposal

B1. bool Globally_Ordered_Ready(int seq)
B2.   if Global_History[seq] contains a Proposal and \floor{N/2} Accepts from the same view
B3.      return TRUE
B4.   else
B5.      return FALSE

C1. Advance Aru()
C2.   i $\leftarrow$ Local_Aru +1
C3.   while (1)
C4.      if Global_History[i].Ordered is not empty
C5.         Local_Aru++
C6.         i++
C7.      else
C8.         return
\end{Verbatim}
  \caption{Global Ordering Utility Procedures}
  \label{11-order-util}
\end{figure}

\begin{figure}[htbp]
  \centering
\scriptsize
\begin{Verbatim}[commandchars=\\\{\},codes={\catcode`$=3},frame=single,baselinestretch=.85]
Client_Update_Handler(Client_Update U):
A1.   if(State = LEADER_ELECTION)
A2.      if(U.server_id != My_server_id)
A3.         return
A4.      if(Enqueue_Update(U))
A5.         Add_to_Pending_Updates(U)
A6.   if(State = REG_NONLEADER)
A7.      if(U.server_id = My_server_id)
A8.      Add_to_Pending_Updates(U)
A9.      SEND to leader: U
A10.  if(State = REG_LEADER)
A11.     if(Enqueue_Update(U))
A12.        if U.server_id = My_server_id
A13.           Add_to_Pending_Updates(U)
A14.     Send_Proposal()

B1. Upon expiration of Update_Timer(client_id):
B2.   Restart Update_Timer(client_id)
B3.   if(State = REG_NONLEADER)
B4.      SEND to leader: Pending_Updates[client_id]
\end{Verbatim}
  \caption{Client Update Handling}
  \label{12-client}
\end{figure}

\begin{figure}[htbp]
  \centering
\scriptsize
\begin{Verbatim}[commandchars=\\\{\},codes={\catcode`$=3},frame=single,baselinestretch=.85]
A1. bool Enqueue_Update(Client_Update U)
A2.   if(U.timestamp $\leq$ Last_Executed[U.client_id])
A3.      return false
A4.   if(U.timestamp $\leq$ Last_Enqueued[U.client_id])
A5.      return false
A6.   Add U to Update_Queue
A7.   Last_Enqueued[U.client_id] $\leftarrow$ U.timestamp
A8.   return true

B1. Add_to_Pending_Updates(Client_Update U)
B2.   Pending_Updates[U.client_id] $\leftarrow$ U
B3.   Set Update_Timer(U.client_id)
B4.   **Sync to disk

C1. Enqueue_Unbound_Pending_Updates()
C2.   For each Client_Update U in Pending_Updates[]
C3.      if U is not bound and U is not in Update_Queue
C4.         Enqueue_Update(U)

D1. Remove_Bound_Updates_From_Queue()
D2.   For each Client_Update U in Update_Queue
D3.      if U is bound or U.timestamp $\leq$ Last_Executed[U.client_id] or
            (U.timestamp $\leq$ Last_Enqueued[U.client_id] and U.server_id $\neq$ My_server_id)
D4.         Remove U from Update_Queue
D5.         if U.timestamp > Last_Enqueued[U.client_id]
D6.            Last_Enqueued[U.client_id] $\leftarrow$ U.timestamp
\end{Verbatim}
  \caption{Client Update Handling Utility Functions}
  \label{13-client-util}
\end{figure}

\begin{figure}[htbp]
  \centering
\scriptsize
\begin{Verbatim}[commandchars=\\\{\},codes={\catcode`$=3},frame=single,baselinestretch=.85]
Recovery:
A1.  if no data exist on stable storage /* Initialization */
A2.    Create stable storage files
A3.    Last_Attempted $\leftarrow$ 0
A4.    Last_Installed $\leftarrow$ 0
A5.    Local_Aru $\leftarrow$ 0
A6.    Last_Proposed $\leftarrow$ 0
A7.    Progress_Timer $\leftarrow$ default timer value
A8.    **Sync to disk
A9.  else
A10.   Read State from stable storage
A11.   Read Last_Attempted and Last_Installed from stable storage
A12.   Rebuild Global_History from message log
A13.   For each pending update U
A14.      Add_to_Pending_Updates(U)
A15.   Compute Local_Aru from Global_History[]
A16. Shift_to_Leader_Election(Last_Attempted + 1)
\end{Verbatim}
  \caption{Recovery Procedure}
  \label{14-recover}
\end{figure}


\clearpage

\mysec{Paxos for System Builders in DistAlgo}
\label{app-orig-da}
This contains the complete DistAlgo program for Paxos-SB.
\lstinputlisting[style=dacodestyle,numbers=left,basewidth = {.43em}]{\src}

\mysec{A Run of Paxos for System Builders in DistAlgo}
\label{app-run}
The following shows a run of the DistAlgo program with 3 servers with 3
clients and 100,000 requests each client.
{
\scriptsize
\begin{verbatim}
> python -m da .paxosSB_orig.da
[110] da.api<MainProcess>:INFO: <Node_:3a001> initialized at 127.0.0.1:(UdpTransport=42511, TcpTransport=44017).
[110] da.api<MainProcess>:INFO: Starting program <module 'paxosSB_orig from '.\paxosSB_orig.da'>...
[110] da.api<MainProcess>:INFO: Running iteration 1 ...
[110] da.api<MainProcess>:INFO: Waiting for remaining child processes to terminate...(Press "Ctrl-Brk" to force kill)
[45] paxosSB_orig.Client<Client:de006>:OUTPUT: ________ Client id:  1
[124] paxosSB_orig.Client<Client:de005>:OUTPUT: ________ Client id:  0
======== Average latency for 100000 requests:  0.0016162170124053956 seconds. Total time:  162.07355737686157
======== Average latency for 100000 requests:  0.0016252124166488647 seconds. Total time:  162.86385893821716
[165426] da.api<MainProcess>:INFO: Main process terminated.
\end{verbatim}
}

\end{document}